\documentclass[aps,prl,twocolumn,floats,epsf,preprintnumbers,superscriptaddress,nofootinbib,amsmath]{revtex4}

\usepackage{graphicx}
\usepackage{amssymb,amsmath}
\usepackage{mathrsfs}
\usepackage{slashed}
\usepackage{indentfirst}
\usepackage{graphicx}
\usepackage{xcolor}
\usepackage{dsfont}
\usepackage{ulem}

\usepackage[colorlinks,
            linkcolor=black,
            anchorcolor=black,
            citecolor=black
            ]{hyperref}

\begin{document}

\preprint{NCTS-PH/1721}


\title{Radiative corrections to Higgs couplings with weak gauge bosons in custodial multi-Higgs models}

\author{Cheng-Wei Chiang}
\affiliation{Department of Physics, National Taiwan University, Taipei, Taiwan 10617, R.O.C.}
\affiliation{Institute of Physics, Academia Sinica, Taipei, Taiwan 11529, R.O.C.}
\affiliation{Physics Division, National Center for Theoretical Sciences, Hsinchu, Taiwan 30013, R.O.C.}
\affiliation{Kavli IPMU, University of Tokyo, Kashiwa, 277-8583, Japan}

\author{An-Li Kuo}
\affiliation{Department of Physics, National Central University, Chungli, Taiwan 32001, R.O.C.}

\author{Kei Yagyu}
\affiliation{INFN, Sezione di Firenze, and Department of Physics and Astronomy, 
University of Florence, Via G. Sansone 1, 50019 Sesto Fiorentino, Italy}


\begin{abstract}

We calculate 1-loop radiative corrections to the $hZZ$ and $hWW$ couplings in models with next--to--simplest Higgs sectors satisfying the electroweak $\rho$ parameter equal to 1 at tree level: the Higgs singlet model, the two-Higgs doublet models, and the Georgi-Machacek model.  Under theoretical and current experimental constraints, the three models have different correlations between the deviations in the $hZZ$ and $hWW$ couplings from the standard model predictions.  In particular, we find for each model predictions with no overlap with the other two models. 
\end{abstract}

\pacs{}

\maketitle


{\it Introduction---}
The particle spectrum of the Standard Model (SM) has been completed by the discovery of a Higgs boson ($h$) at the LHC, of which properties are found to be consistent with SM expectations within uncertainties. 
However, it is widely accepted that the SM should be replaced by a more fundamental theory 
at a higher energy scale to address such issues as the origin of neutrino masses and the existence of dark matter among others.  A pressing task in particle physics is to search for new particles in or footprints left by such new physics (NP) in current and future experiments.

Since many NP models predict a non-minimal structure in their Higgs sector, determining its true shape turns out to be an important probe of physics beyond the SM. 
If we discover a second Higgs boson in future collider experiments, this must be direct evidence of a non-minimal Higgs sector. 
In that case, by measuring its properties, such as mass, width and cross sections, 
we can reconstruct the structure of the Higgs sector. 
Alternatively, a non-minimal Higgs sector can be explored indirectly by measuring the $h$ couplings to other SM particles including itself. If non-zero deviations in $h$ couplings from the SM predictions 
are found with a specific pattern, we can extract information of the Higgs sector 
such as the representation of extra scalar fields by comparing measured values with theory predictions~\cite{fingerprint}.  
Therefore, albeit indirect, a precision determination of the $h$ couplings is an effective probe of NP,  
particularly when new Higgs bosons are beyond the reach of available colliders.

As experimental precision becomes higher, we are forced to go beyond the tree-level calculation for the 
$h$ couplings in order to make a sensible comparison.  
In particular, the $hZZ$ and $hWW$ couplings will be measured with higher precisions (better
than 1\% level) than the others at future $e^+e^-$ colliders, {\it e.g.}, the International Linear Collider (ILC)~\cite{ILC}. 
Therefore, precise calculations of these couplings by taking into account electroweak radiative corrections 
are inevitable for future comparisons.

In this Letter, we study deviations in the $hZZ$ and $hWW$ couplings from their SM predictions at 1-loop level in models having a non-minimal Higgs sector with electroweak $\rho$ parameter equal to 1 at tree level ($\rho_{\rm tree} = 1$). 
In such models, the coupling relation $g_{hWW} = g_{hZZ} \cos^2\theta_W$ with $\theta_W$ being the weak mixing angle is satisfied at tree level. 
However, this is no longer valid in general when loop effects are taken into account, as they receive different radiative corrections.  
We will show different characteristic deviations in these couplings at 1-loop level for different models.

{\it Renormalized $hVV$ vertex---}
First of all, we define the renormalized $hV^\mu V^\nu$ vertices ($V = Z,W$) in the $SU(2)_L \times U(1)_Y$ gauge theory, in which the Higgs sector is assumed to have at least one $SU(2)_L$ doublet field. 
In general, the renormalized $hV^\mu V^\nu$ vertices can be decomposed in terms of three form factors as follows:
\begin{align}
\hspace{-3mm}
\hat{\Gamma}_{hVV}^{\mu\nu} 
= \hat{\Gamma}_{hVV}^1 g^{\mu\nu}
+\hat{\Gamma}_{hVV}^2 p_1^\nu p_2^\mu 
+\hat{\Gamma}_{hVV}^3 \epsilon^{\mu\nu\rho\sigma}p_{1\rho} p_{2\sigma}
~, 
\end{align}
where $p_1^\mu$ and $p_2^\mu$ are the incoming 4-momenta of the gauge bosons, and $\hat{\Gamma}_{hVV}^{2,3}$ only appear from 1-loop 1-particle irreducible (1PI) diagram contributions.  
On the other hand, $\hat{\Gamma}_{hVV}^1$ comprises three parts:
\begin{align}
\hspace{-3mm}
\hat{\Gamma}_{hVV}^{1} 
= \frac{2m_V^2}{v}\kappa_V^{} + \Gamma_{hVV}^{\text{1PI}}
+ \delta \Gamma_{hVV}
~, \label{gam1}
\end{align}
where $m_V$ is the gauge boson mass, $\kappa_V^{}$ is the scaling factor from the SM value at tree level, 
and $v\simeq 246$~GeV is the vacuum expectation value (VEV).
The three terms represent respectively contributions from the tree-level coupling, 1PI diagrams, and the counterterm.  We are interested in the scaling factor at 1-loop level defined by
\begin{align}
\hat{\kappa}_V^{}(p^2) 
\equiv 
\frac{\hat{\Gamma}_{hVV}^1(m_V^2,p^2,m_h^2)_{\text{NP}}}{\hat{\Gamma}_{hVV}^1(m_V^2,p^2,m_h^2)_{\text{SM}}}
~, \label{hatk}
\end{align}
where we have noted the momentum dependence of one of the gauge bosons as the couplings are expected to be measured primarily through the $Vh$ production at future colliders.  For later convenience, we also introduce
\begin{align}
 \Delta  \hat{\kappa}_V^{}(p^2) 
 \equiv 
 \hat{\kappa}_Z(p^2) - \hat{\kappa}_W(p^2) ~.  \label{delk}
\end{align}

\begin{table*}[t!]
\begin{ruledtabular}
{\small
\begin{tabular}{c c c c c c c c}
         & Extra fields & $v^2$                  & $\tan\beta$                   & $\kappa_V^{}$    & $\kappa_V^H$ & $\kappa_Z^{H_5}(\kappa_W^{H_5})$ & $\delta \kappa_V^{}$  \\\hline
HSM      &$S({\bf 0},0)$& $v_\Phi^2$              & --                            & $c_\alpha$       & $s_\alpha $   & -- &$-s_\alpha \delta \alpha$                     
\\
2HDMs    &$\Phi'({\bf 1/2},1/2)$& $v_\Phi^2+v_{\Phi'}^2$   & $v_\Phi/v_{\Phi'}$ & $s_{\beta-\alpha}$ & $c_{\beta-\alpha}$ & --  &$c_{\beta-\alpha}(\delta \beta - \delta \alpha )$  
\\
GM model &$\chi({\bf 1},1),\xi({\bf 1},0)$& $v_\Phi^2+8v_{\Delta}^2$ & $v_\Phi/(2\sqrt{2}v_{\Delta}^{})$& $c_\alpha s_\beta -\frac{2\sqrt{6}}{3}s_\alpha c_\beta  $ & 
$s_\alpha s_\beta +\frac{2\sqrt{6}}{3}c_\alpha c_\beta  $ & $\frac{2c_\beta}{\sqrt{3}}$~$(-\frac{c_\beta}{\sqrt{3}})$   
& $\frac{\partial \kappa_V^{}}{\partial \alpha}\delta \alpha + \frac{\partial \kappa_V^{}}{\partial \beta}\delta \beta + C_V \delta \rho$  
\\
\end{tabular}
}
\end{ruledtabular}
\caption{Extra Higgs fields and parameters in the HSM, 2HDMs and GM model with the shorthand notation
$s_\theta \equiv \sin\theta$, $c_\theta \equiv \cos\theta$, and $t_\theta \equiv \tan\theta$. 
Each scalar field is labeled with the quantum numbers $(T,Y)$ of weak isospin and hypercharge, respectively.  For the GM model, $C_Z = \kappa_V/2$ and $C_W = -t_\beta/2 \times (\partial \kappa_V^{}/\partial \beta)$.  
The mixing angle $\alpha$ is defined in the main text.}
\label{prop}
\end{table*}

Assuming that $h$ mixes with extra neutral Higgs bosons generically denoted by $\phi$ at tree level via mixing angles $\alpha_{\phi h}$, then we generally have
\begin{align}
\delta \Gamma_{hVV} 
&= \frac{2m_V^2}{v}\Bigg[
\kappa_V^{} \left(
\frac{\delta m_V^2}{m_V^2} - \frac{\delta v}{v} + \delta Z_V + \frac{1}{2}\delta Z_h 
\right)
\notag\\
&~~~~~~~~~~~ + \delta \kappa_V^{} + \sum_\phi \kappa_V^\phi (\delta \alpha_{\phi h} + \delta Z_{\phi h})
\Bigg]
~, \label{del_gam}
\end{align}
where $\delta m_V^2$, $\delta v$ and $\delta \kappa_V^{}$ denote the counterterms for $m_V^2$, $v$ and $\kappa_V^{}$, respectively, 
while $\delta Z_V$ ($\delta Z_h$) is the field renormalization factor for $V^\mu$ ($h$). 
The last term appears due to the off-diagonal element of the field shift for $\phi$; namely, 
$\phi \to \delta \tilde{Z}_{\phi h}h +\cdots = (\delta \alpha_{\phi h} + \delta Z_{\phi h})h+\cdots$, 
with $\delta \alpha_{\phi h}$ being the counterterm of $\alpha_{\phi h}$. 
The factor $\kappa_V^\phi$ is the $\phi VV$ coupling normalized to the SM $hVV$ coupling at tree level. 
We note that $\delta \kappa_V^{} = \kappa_V^\phi = 0$ in the SM . 
The counterterm $\delta v$ can be rewritten by employing the tree-level relation, $v^2 = m_W^2 s_W^2/(\pi \alpha_{\text{em}})$, as follows: 
\begin{align}
\frac{\delta v}{v} 
= \frac{1}{2}\left(
\frac{\delta m_W^2}{m_W^2} + \frac{\delta s_W^2}{s_W^2} 
- \frac{\delta \alpha_{\text{em}}}{\alpha_{\text{em}}} 
\right)
~, 
\end{align}
where $\alpha_{\text{em}}$ is the fine-structure constant, $s_W = \sin\theta_W$, and 
\begin{align}
\delta s_W^2 =
-\frac{m_W^2}{m_Z^2\rho_{\text{tree}}}
\left(
\frac{\delta m_W^2}{m_W^2}-\frac{\delta m_Z^2}{m_Z^2}
-\frac{\delta \rho}{\rho_{\text{tree}}} 
\right)
~,  
\end{align}
with $\delta \rho$ being the counterterm of the $\rho$ parameter. 

In models with a Higgs sector composed only of $SU(2)_L$ singlet and/or doublet fields, $\rho_{\text{tree}} =1$ identically and $\delta \rho$ does not appear. 
In contrast, models with higher $SU(2)_L$ representations such as Higgs triplets, $\rho_{\text{tree}} \not= 1$ in general. 
In such a model, $\delta \rho$ appears as an additional degree of freedom in comparison with the former models, 
and hence we need an additional condition to determine it. 
We can impose the following condition by taking the electroweak oblique parameter $T$~\cite{Peskin} as {\it input} as follows:
\begin{align}
\alpha_{\text{em}}T &= 
\frac{\Pi_{ZZ}^{\text{1PI}}(0)-\Pi_{ZZ}^{\text{1PI}}(0)\big|_{\text{SM}}}{m_Z^2}
-\frac{\Pi_{WW}^{\text{1PI}}(0)-\Pi_{WW}^{\text{1PI}}(0)\big|_{\text{SM}}}{m_W^2} \notag\\
&+\delta \rho,\label{delt}
\end{align}
where $\Pi_{XY}^{\text{1PI}}$ are the 1PI diagram contributions to the 2-point functions of particles $X$ and $Y$.  
For gauge bosons, $\Pi_{XY}^{\text{1PI}}$ are defined by their transverse components.
From Eq.~(\ref{delt}), $\delta \rho$ is determined. 

The other counterterms can be determined by imposing the usual on-shell renormalization conditions~\cite{OS1,OS2,Hollik,Denner}:
\begin{align}
\begin{split}
\delta m_V^2 
&= \Pi_{VV}^{\text{1PI}}(m_V^2)
~, \\
\frac{\delta\alpha_{\text{em}}}{\alpha_{\text{em}}} 
&= \frac{d}{dp^2}\Pi_{\gamma\gamma}^{\text{1PI}}(p^2)\Big|_{p^2=0}
~, \\
\delta Z_W 
&=  -\frac{\delta\alpha_{\text{em}}}{\alpha_{\text{em}}} +\frac{2c_W^{}}{s_W^{}}\frac{\Pi_{ Z\gamma}^{\text{1PI}}(0)}{m_Z^2}
+\frac{\delta s_W^2}{s_W^2}
~, \\
\delta Z_Z 
&= \delta Z_W -\frac{2s_W^{}}{c_W^{}}\frac{\Pi_{ Z\gamma}^{\text{1PI}}(0)}{m_Z^2} - \frac{\delta s_W^2}{c_W^2}
~, \\
\delta Z_h 
&= \frac{d}{dp^2}\Pi_{hh}^{\text{1PI}}(p^2)\Big|_{p^2=m_h^2}
~,\\
\delta Z_{\phi h} 
&= \frac{1}{m_\phi^2 - m_h^2}\left[\Pi_{\phi h}^{\text{1PI}}(m_h^2)-\Pi_{\phi h}^{\text{1PI}}(m_\phi^2) \right]
~. \label{del_ij}
\end{split}
\end{align}
Although $\delta \alpha_{\phi h}$ is also given in terms of $\Pi_{\phi h}^{\text{1PI}}$, 
this is eventually cancelled by the $\delta \alpha_{\phi h}$ term coming from $\delta \kappa_V^{}$ 
in the concrete models considered in this work. 
The expression of $\delta \kappa_V^{}$ in the concrete models is presented in Table~\ref{prop}.

{\it Concrete models---}
We consider three models with next-to-simplest Higgs sectors satisfying $\rho_{\text{tree}}=1$: Higgs singlet model (HSM), two-Higgs doublet models (2HDMs)~\cite{Branco:2011iw} and Georgi-Machacek (GM) model~\cite{Georgi:1985nv,Chanowitz:1985ug}.  In each model, the Higgs sector comprises one $SU(2)_L$ doublet $\Phi$ with hypercharge $Y=1/2$ and additional scalar multiplet fields given in Table~\ref{prop}. 
In the HSM, the singlet VEV $v_S^{}$ does not contribute to the electroweak symmetry breaking and fermion masses. 
In addition, the value of $v_S^{}$ can be arbitrary without changing physics, so
that we simply take $v_S^{}=0$. 
In the 2HDMs, we impose a softly-broken $Z_2$ symmetry to forbid tree-level flavor-changing neutral currents, and assume CP conservation for simplicity.
Finally, we assume in the GM model that the triplet fields have the same VEV 
($v_\chi = v_\xi \equiv v_\Delta^{}$) so that $\rho_{\text{tree}}=1$.  
We note that $\delta \rho$ appears in $\delta \Gamma_{hVV}$ only in the GM model, which can be expressed as 
$\delta \rho = 8v_\Delta^{}\delta \nu/v^2 $, with $\nu \equiv v_\chi -v_\xi$ and $\nu = 0 + \delta \nu$.
In these three models, the sum rule $v$ and $\beta$ parameter defined by $s_\beta = v_\Phi/v$ are given in Table~\ref{prop}.

There are two CP-even Higgs bosons in the HSM and the 2HDMs, while three in the GM model. 
The relation between the weak eigenstates and the mass eigenstates is given by the following orthogonal transformations:
\small
\begin{align}
&\hspace{-0.2cm}\begin{pmatrix}
S \\
\Phi_r^0
\end{pmatrix} 
\hspace{-2pt}
= 
\hspace{-2pt}
\begin{pmatrix}
c_{\alpha} & -s_\alpha \\
s_\alpha & c_\alpha
\end{pmatrix}
\hspace{-2pt}
\begin{pmatrix}
H\\
h
\end{pmatrix}
,~
\begin{pmatrix}
\Phi_r^{\prime 0}\notag\\
\Phi_r^0
\end{pmatrix} 
\hspace{-2pt}
= 
\hspace{-2pt}
\begin{pmatrix}
c_{\alpha} & -s_\alpha \\
s_\alpha & c_\alpha
\end{pmatrix}
\hspace{-2pt}
\begin{pmatrix}
H\\
h
\end{pmatrix}
,
\\ 
&\hspace{-0.2cm} 
\begin{pmatrix}
\xi^0\\
\Phi_r^0\\
\chi_r^0
\end{pmatrix}
 = \begin{pmatrix}
\frac{1}{\sqrt{3}} &0& -\sqrt{\frac{2}{3}}\\
0 & 1 &0\\
\sqrt{\frac{2}{3}} & 0 & \frac{1}{\sqrt{3}}
\end{pmatrix}
\begin{pmatrix}
c_\alpha & -s_\alpha & 0 \\
s_\alpha & c_\alpha & 0 \\
0 & 0 & 1 
\end{pmatrix}
\begin{pmatrix}
H\\
h \\
H_5
\end{pmatrix}, \notag
\end{align}
\normalsize
where $\Phi_r^{0} = \sqrt{2}\text{Re}\,\Phi^{0}$, 
$\Phi_r^{\prime 0} = \sqrt{2}\text{Re}\,\Phi^{\prime 0}$ and $\chi_r^{0} = \sqrt{2}\text{Re}\,\chi^0$. 
Using the above relations, we find $\delta \alpha_{Hh} = \delta \alpha$ in all the three models 
and $\delta \alpha_{H_5h} = 0$ in the GM model. 

In the 2HDMs and the GM model, $\delta \beta$ appears in $\delta \kappa_V^{}$. 
This can be determined by demanding the condition:
\begin{align}
\hat{\Pi}_{G^0{\cal A}}(0) = \hat{\Pi}_{G^0{\cal A}}(m_{\cal A}^2) = 0
~, \label{del_beta1}
\end{align}
where $\hat{\Pi}_{G^0{\cal A}}$ is the renormalized mixed 2-point function between the neutral Nambu-Goldstone boson ($G^0$) 
absorbed into $Z_L$ and the physical CP-odd Higgs boson ${\cal A}$ with mass $m_{\cal A}$. 
From Eq.~(\ref{del_beta1}), we obtain 
\begin{align}
\delta \beta 
= -\frac{1}{2m_{\cal A}^2}\left[
\Pi_{G^0{\cal A}}^{\text{1PI}}(0)+\Pi_{G^0{\cal A}}^{\text{1PI}}(m_{\cal A}^2)
\right]
~. \label{del_beta2}
\end{align}
It is known that there remains gauge dependence in $\delta \beta$ determined from Eq.~\eqref{del_beta1}~\cite{Freitas}.
Such a dependence can be removed by use of the so-called pinch technique~\cite{Cornwall:1981zr}, where additional 
gauge-dependent pinch terms extracted from vertex corrections and box diagrams in 
a $f\bar{f} \to f\bar{f}$ scattering process are added to the mixed 2-point function.  
For the 2HDMs, the gauge-invariant renormalization scheme using the pinch technique 
has been discussed in Refs.~\cite{Santos,Yagyu}. 
For the GM model, we can analogously apply the pinch technique to define the gauge-invariant $\delta \beta$. 
Nevertheless, a novel difference is that the gauge dependence is not cancelled out within the $G^0$--${\cal A}$ mixing diagram even after adding the pinch terms, for its gauge-dependent part from the custodial $SU(2)$ 5-plet Higgs boson loop diagrams does not vanish 
(note that the 5-plet Higgs bosons do not couple to fermions, and hence do not contribute to the pinch terms). 
In fact, the cancellation occurs among the $G^0$--${\cal A}$, $G^0$--$G^0$ and ${\cal A}$--${\cal A}$ diagrams.  A detailed discussion about this gauge dependence issue in the GM model will be presented in a separate work~\cite{future}, where the full analytic expressions needed to compute the renormalized $hV^\mu V^\nu$ vertices are given.  
Here we simply adopt the renormalization condition given in Eq.~(\ref{del_beta1}), since the main focus is 
to see the difference in $\hat{\kappa}_Z^{}$ and $\hat{\kappa}_W^{}$ among the three models, where 
finite corrections to maintain gauge invariance is expected to be negligibly small as shown in Ref.~\cite{Yagyu} for the 2HDMs.  For the detailed analytic expressions of various 1PI diagram contributions, 
see Ref.~\cite{rad_hsm} for the HSM and Ref.~\cite{rad_2hdm} for the 2HDMs. 

{\it Numerical results---}
We use the following SM inputs~\cite{PDG} $(m_Z^{},\,m_t,\,m_b,\,m_c,\,m_\tau,\,m_h)=(91.1875,\,173.21,\,4.66,\,1.275,\,1.77684,\,125)~\text{GeV}$, 
$G_F=1.16639\times 10^{-5}$ GeV$^{-2}$, $\alpha_{\text{em}}^{-1}=137.035989$ and $\Delta\alpha_{\text{em}}=0.06635$ as the shift in $\alpha_{\text{em}}$ from zero energy to $m_Z^{}$. 
As a reference, the 1-loop corrections to the $hZZ$ and $hWW$ couplings 
in the SM are respectively about $-1.2~(+1.0)\%$  and $+0.4~(1.3)\%$ for $\sqrt{p^2}=250$ (500) GeV with respect to their tree-level values. 
Moreover, these radiative corrections increase monotonically with $\sqrt{p^2}$.

For the numerical analysis of $\hat{\kappa}_Z^{}$ and $\Delta\hat{\kappa}_V^{}$ defined 
in Eqs.~(\ref{hatk}) and (\ref{delk}) for the three models, we scan model parameters allowed under two types of constraints. We define a Type-A constraint as imposing the requirements of vacuum stability~\cite{pu_HSM,vs_THDM,Hartling:2014zca}, perturbative unitarity~\cite{pu_HSM,pu_THDM,Aoki:2007ah}, and the oblique $S$ and $T$ parameters~\cite{Peskin}, and a Type-B constraint as further (after the Type-A constraint) considering the Higgs signal strengths compiled in Ref.~\cite{data} and direct searches for extra Higgs bosons, all at $95\%$ confidence level.  
For the direct searches of the 2HDMs, we employ the neutral Higgs boson searches done in Refs.~\cite{2hdm-direct}.  
We note that the constraint from the search for charged Higgs bosons via the $tb$ decay is less stringent as compared to that from the neutral Higgs boson searches.  For example, there is no constraint on the charged Higgs boson mass for $\tan\beta > 1$~\cite{2hdm-direct5}. 
For the GM model, we use the search for doubly-charged Higgs bosons via the same-sign diboson decay reported in Ref.~\cite{directsearch}.
As seen in Eq.~(\ref{delt}), 
$T$ is a free parameter in the GM model unlike the HSM and the 2HDMs and will be fixed at 0 for illustration purposes.  
Moreover, radiative corrections to $hVV$ couplings have a linear dependence on the $T$ parameter.

\begin{figure}[th]
\centering
\includegraphics[scale=0.2]{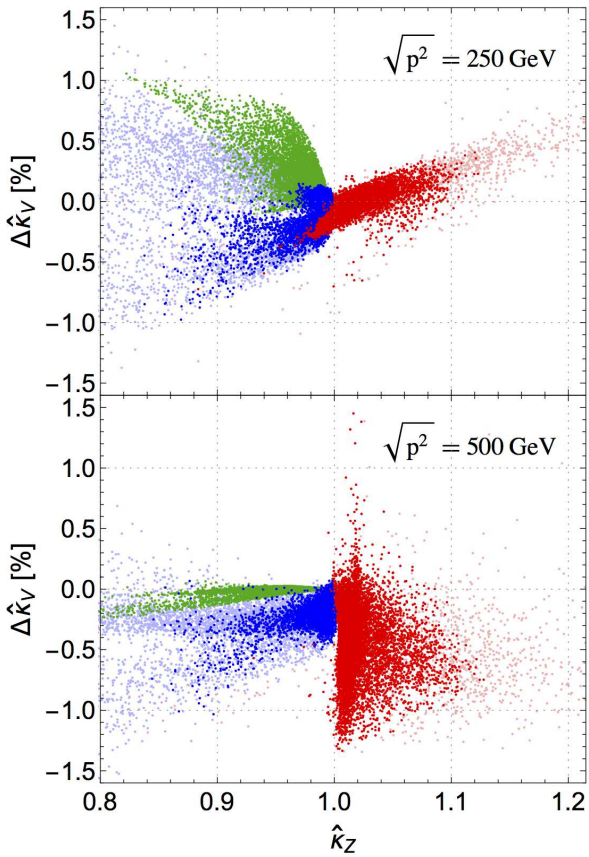}
\caption{Correlation between $\hat\kappa_Z$ and $\Delta\hat\kappa_V$ for $\sqrt{p^2}=250$~GeV (upper plot) 
and 500~GeV (lower plot) in the HSM (green), Type-I 2HDM (blue) and GM model (red).  
The light (dark) dots are allowed under Type-A (-B) constraint.
Model parameters are scanned or fixed as follows: 
$0.3\leq m_H \leq 1$ TeV, $0.8 \leq c_\alpha\leq 1$, $0\leq \lambda_{\Phi S}\leq 10$, $\lambda_S=\mu_S=0$ for the HSM; 
$0.3\leq m_{H^\pm},m_A,m_H \leq 1$ TeV, $0.8\leq s_{\beta-\alpha}\leq 1$, $0\leq M\leq 1$ TeV, $1.5 \le t_\beta \le 10$ for the 2HDM; and
$-0.628\leq\lambda_2\leq1.57$, $-1.57\leq\lambda_3\leq1.88$, $\left|\lambda_4\right|\leq2.09$, $\left|\lambda_5\right|\leq8.38$, $-650\leq\mu_1\leq0$ GeV, $-400\leq\mu_2\leq50$ GeV, $180\leq m_2\leq450$ GeV for the GM model. For the notation of scanned parameters, see Ref.~\cite{Yagyu} for the HSM and 2HDMs and Ref.~\cite{GM} for the GM model.}
\label{kztodk}
\end{figure}

Correlations between $\hat\kappa_Z$ and $\Delta\hat\kappa_V$ are shown in Fig.~\ref{kztodk} 
for the HSM (green), Type-I 2HDM (blue) and GM model (red) under the Type-A (lighter colors) 
and Type-B (darker colors) constraints. 
The upper (lower) plot is for $\sqrt{p^2}=250$~GeV (500~GeV).
We note that for the other three types (Type-II, -X and -Y) of Yukawa interactions in the 2HDMs, 
the results under the Type-A constraint are almost the same as that in the Type-I 2HDM while those under the Type-B constraint can be drastically different, with smaller allowed regions on the $\hat{\kappa}_Z$--$\Delta \kappa_V$ plane as compared to the Type-I 2HDM.

By comparing the results under Type-A and Type-B constraints, one can see the impact of LHC Higgs data.  
One obvious feature is that the three models have distinctively different distributions in the plots, 
and can be used to distinguish among the models when the $hVV$ couplings are measured to sufficiently high precision.  
In the HSM and Type-I 2HDM, both allowed dots under the Type-A and Type-B constraints have distributions with $\hat\kappa_V \alt 1$.  
For the GM model, however, most Type-B points have $\hat\kappa_V \agt 1$.  
For the three models in the case of $\sqrt{p^2}=250$~GeV, 
the maximum $|\Delta\hat\kappa_V|$ occurs when $\hat\kappa_Z$ has its largest deviation from unity.  
In the case of $\sqrt{p^2}=500$~GeV, however, largest $|\Delta\hat\kappa_V|$ occurs when $\hat\kappa_Z\simeq 1$.  
Moreover, the possible range of $\Delta\hat\kappa_V$ for the GM model in the latter case is significantly larger than the former 
in contrast to HSM.  
Given the theoretical and current experimental constraints, the GM model has the largest 
allowed $\left|\Delta\hat\kappa_V\right|$ among the models considered here.  

{\it Summary---}
We have computed 1-loop renormalized $hVV$ couplings for the three models with next-to-simplest Higgs sectors giving $\rho_{\text{tree}} = 1$, and found that they have distinctively different distributions under the theoretical and current experimental constraints.  In particular, $\Delta\hat\kappa_V$ sometimes is large enough to be measurable at future $e^+e^-$ colliders such as the ILC, and can serve as an additional means to discriminate models.  We also note that gauge dependence in the GM model is cancelled using the pinch technique in a nontrivial way.

{\it Acknowledgments---}
This research was supported in part by the Ministry of Science and Technology of Taiwan under Grant No.\ MOST 104-2628-M-002-014-MY4.

\vspace{-4mm}

\end{document}